\documentclass[12pt]{iopart}

\begin{document}

\title
{Alternative commutation relations, star-products
and tomography}

\author{Olga V  Man'ko,$^\dagger$
V I Man'ko$^\dagger$
and G Marmo$^\ddagger$}
\address{
${}^\dagger$P. N. Lebedev Physical Institute,
Leninskii Prospect 53, Moscow 119991 Russia\\
${}^\ddagger$Dipartimento di Scienze Fisiche, 
Universit\`a ``Federico II'' di Napoli and
Istituto Nazionale di Fisica Nucleare, Sezione di
Napoli, Complesso Universitario di Monte S.~Angelo,
Via Cintia, I-80126 Napoli, Italy}

E-mail: 

omanko@sci.lebedev.ru

manko@sci.lebedev.ru

gimarmo@na.infn.it

\begin{abstract}
Invertible maps from operators of quantum obvservables 
onto functions of $c$-number arguments and their associative 
products are first assessed.
Different types of maps like Weyl--Wigner--Stratonovich map
and $s$-ordered quasidistribution are discussed.
The recently introduced symplectic tomography map of 
observables (tomograms)
related to the Heisenberg--Weyl group is shown to belong
to the standard framework of the maps from quantum observables
onto the  $c$-number functions.
The star-product for symbols of the quantum-observable for each
one of the maps (including the tomographic map) and explicit 
relations among different star-products are obtained. 
Deformations of the Moyal star-product and alternative 
commutation relations are also considered.
\end{abstract}

{\bf keywords}: star-product, tomography, quasidistributions

\pacs{0365T, 0365S}


\section{Introduction}
 
In a two-pages paper~\cite{Wigner51}, fifty years ago, Wigner 
raised the question of the uniqueness of the commutation relations 
compatible with the evolution of a quantum oscillator.
Several papers have been devoted to this problem
since (see~\cite{Ramon} and references therein). 
More recently, in connection with the problem of integrability, the
problem has received new attention. Indeed, it is well known that
alternative and compatible Poisson brackets appear in connection 
with the problem of complete integrability within a classical 
framework~\cite{Ibort}.
On the other hand, classical mechanics may be derived, in some appropriate
limit, from quantum mechanics. It is then a natural question to ask which
alternative quantum structures, after taking the `classical limit',
would reproduce the alternative known Hamiltonian descriptions.
This paper belongs to this set of ideas even though it will not be
concerned with the problem of complete integrability.
We shall concentrate our attention on the alternative structures 
within the framework of Heisenberg picture for operators acting on 
Hilbert spaces of infinite and finite dimensions. We shall also make 
considerations on a generalized version of Ehrenfest picture.
Recently, the activity connected with quantum computing has regained a
great interest for finite-level quantum systems, in addition, within 
this framework, one does not have to worry about domain problems 
for operators, therefore we shall indulge a little also with finite 
level quantum systems. Having in mind the comparison with the classical 
limit, a predominant role will be played by the Wigner map~\cite{Wigner32},
associating functions on the phase space with operators acting on the 
space of states. In this connection, we shall also consider tomographic 
descriptions of Wigner functions~\cite{ManciniPhysLett,ManciniFoundPhys}
 and show how they behave with respect 
to alternative products.
If one considers the behaviour of star-products with respect to the
`deformation' parameter $\hbar$, one shows that 
$s$-quasidistributions~\cite{CahillGlauber67}
give rise to star-products, which are different only at the order of  
$\hbar^2$ and onwards, but coincide at the order of $\hbar$.

The formalisms of quantum and classical mechanics are 
drastically different in
the sense that the physical observables of 
classical mechanics are described by
$c$-numbers and the quantum observables are described 
by operators acting on
Hilbert space of quantum states~\cite{Dirac}
and the quantum states are associated with 
density matrix for mixed states~\cite{vonNeuman} and wavefunction for 
pure states~\cite{Schroedinger26}.
Due to Heisenberg uncertainty relation~\cite{Heisenberg27},
the existence of conventional joint probability 
distribution on the phase space
is impossible in quantum mechanics.
 
The Wigner function turns out to 
be Weyl symbol of the density operator
and the evolution equation for the 
Wigner function, introduced by Moyal~\cite{Moyal49},
is just a famous example of the possibility to 
formulate quantum mechanics using an
invertible map between density operators and 
functions on the phase space.
Several different such maps 
from density operators (and other
operators) to $c$-number functions 
(or generalized functions)
on the phase space have been introduced. 
Known types of functions 
are singular Glauber--Sudarshan 
quasidistribution~\cite{Glauber63,Sudarshan63},
nonnegative Husimi quasidistribution~\cite{Husimi40}
and Wigner quasidistribution.

Recently, the tomographic map from density operators 
onto homogeneous 
probability distribution functions (tomograms) 
of one random variable $X$
and two real parameters $\mu$ and $\nu$ has been introduced.
Nowadays, the tomographic map is used to reconstruct the 
quantum state and to obtain the Wigner function by measuring 
the state tomogram.
This map has been used to provide a formulation of 
quantum mechanics~\cite{ManciniPhysLett,ManciniFoundPhys},
in which the quantum state is described by 
conventional nonnegative probability
distribution, alternative to the description of 
the state by the wave function or density operator.
Analogous procedure for a tomographic map for
spin states, i.e., for $SU(2)$-group representations, 
was presented in~[16--18].
The nonredundant spin-tomography scheme was 
suggested by Weigert~\cite{WeigertPRL}.
The relation of the tomographic map for continuous 
position to Heisenberg--Weyl group representations 
has been studied in~\cite{MarmoPhysScripta1}
and the possibility to associate spin tomograms with 
classical linear systems has been presented 
in~\cite{MarmoPhysScripta2}. 
The tomographic map associates operators with functions of 
position measured in a reference frame of the phase space, 
this frame appears as additional `independent variables'
through rotation parameters and scaling parameters, therefore
it is not a map from operators onto functions on 
the phase space.
The tomographic symbols are functions of only one of a pair of
conjugate variables --- position, for instance.
Other two variables are considered as parameters
characterizing the reference frame, namely, 
the rotation and scaling parameters $\theta$
and $\lambda$ related to real parameters 
$\mu$ and $\nu$ as
$\mu=\cos \theta \exp \lambda$, 
$\nu=\sin \theta \exp \,(-\lambda)$.
In the association of operators with 
functions, which are symbols of the
operators, the product of the operators induces 
a special product for symbols which is called the star-product 
of functions. The rigorous mathematical
description of the star-product is presented 
in~\cite{Fedosov,Konzevich}.
Stratonovich has developed~\cite{Stratonovich}
a general approach to construct the map from 
operators onto $c$-number functions and has discussed
quantum systems in terms of the operator symbols.
The approach of~\cite{Stratonovich}
was recently reconsidered in~\cite{BrifMann}
in connection with some tomographic schemes for 
measuring quantum states. Recently
the formula of star-products for several Weyl 
symbols has been given a geometrical flavour 
in~\cite{ZachosJMathPhys}.
A way to generate all Wigner functions has been proposed
in \cite{CUZ},

The star-product of spin-tomograms was investigated 
in a recent paper~\cite{MarmoOlgaPhysScripta}.
One should note that the star-product formalism
for higher spin gauge theories was studied in~\cite{Vas1,Vas2}.
A general consideration of the star-product quantization 
procedure is presented in~\cite{Fronsdal}.
The product of symbols reproduces the associative 
product-rule for the operators. 
There exists a procedure of deformations of the operators, 
e.g., there exist $q$-deformed 
oscillators~\cite{Biedenharn,McF}
related to quantum groups. A physical meaning for 
$q$-oscillators as nonlinear
oscillators with a specific dependence of 
the oscillator frequency on its
amplitude was discussed in~\cite{SolimenoJModPhys}.
The generalization of the deformations taking into 
account other types of 
nonlinearities of vibrations was considered 
in \cite{SudarshanPhysScripta,SudarshanMexico,WVogel}.
Recently, some new deformed associative product of 
operators has been discussed in~\cite{SudarshanJModPhys}
where an additional operator is used.
The dynamics of magnetic dipole was also
studied using the deformed product of spin 
operators~\cite{MarmoOlgaPhysScripta}.
The deformation of products of operators induces 
a deformation of the star-product
of their symbols.
Though the star-product of Weyl symbols is 
well known, the deformations of
the described type of the star-product of Weyl 
symbols have not been studied
(to the best of our knowledge).

The aim of our paper is to present a unified 
approach to
construct both the star-product of symbols 
based on nondeformed 
products of operators and the star-product 
of symbols based on a deformation of 
products.
We show that the tomographic map and the
 formulation of quantum mechanics
in which the state is defined by symplectic 
tomogram can be considered within the framework
of star-product procedure like it was shown 
for spin tomography in \cite{MarmoOlgaPhysScripta}.

A new result is the formula 
for star-product of symbols which
are symplectic tomograms of the operators.
Also we will discuss the deformations
of mentioned symbols in the context of possible 
deformations of the products of
finite and infinite-dimensional matrices. 
Our considerations do not intend to be mathematically 
rigorous, and we assume throughout that various formulae
have meaning when the operators and symbols appearing in 
them are chosen from appropriate spaces.

The paper is organized as follows.

In section~2, a general scheme for associating operators with 
functions and the corresponding star-product construction 
is presented.
In section~3, the example of matrix mechanics is considered.
In section~4, quantum commutators, Poisson brackets and
Heisenberg equations of motion are discussed within the framework
of the general star-product scheme. In section~5, general relations
between different types of maps from operators on functions and 
formulae for intertwining kernels are studied. In section~6, 
the kernel determining star-product of operator-symbols
is discussed and properties
of Weyl symbols are reviewed in section~7. Star-product 
of $s$-ordered symbols is studied in section~8 while star-product
of tomographic symbols is introduced and studied in section~9.
Deformations of star-product are discuused in section~10.
For the aim of completness, in appendix~1 an abstract mathematical
structure of associative products of finite-dimensional vectors and
matrices is elaborated and in appendix~2 an abstract mathematical 
structure of associative product of functions considered as vector 
components is considered.   

\section{General case of functions and operators}

In quantum mechanics, observables are described by 
operators acting on the Hilbert space of states. In order
to consider observables as functions on a phase 
space, we review first a general construction and
provide general relations and properties of a map from
operators onto functions without a concrete realization of the map. 
Given a Hilbert space $H$ and an operator $\hat A $
acting on this space, let us suppose that we have a set
of operators $\hat U({\bf x})$
acting on $H$, a $n$-dimensional
vector ${\bf x}=(x_1,x_2,\ldots,x_n)$ labels the particular
operator in the set. We construct the $c$-number function
$f_{\hat A}({\bf x})$ (we call it the symbol of operator $\hat A$ )
using the definition 
\begin{equation}\label{eq.1}
f_{\hat A}({\bf x})=\mbox{Tr}\left[\hat A\hat U({\bf x})\right].
\end{equation}
For example, the symbol for the Hamiltonian of the free particle 
$$\hat H=\frac{\hat p^2}{2m}$$
reads
$$f_{\hat H}({\bf x})=\frac{1}{2m}\,\mbox{Tr}
\,\Big[\hat p^2\hat U({\bf x})\Big].$$ 
Let us suppose that relation~(\ref{eq.1}) has an inverse,
i.e., there exists a set of operators $\hat D({\bf x})$ acting
on the Hilbert space such that
\begin{equation}\label{eq.2}
\hat A= \int f_{\hat A}({\bf x})\hat D({\bf x})~d{\bf x}
\qquad \mbox{Tr}\,
\hat A= \int f_{\hat A}({\bf x})\,\mbox{Tr}\,
\hat D({\bf x})~d{\bf x}.
\end{equation}
Then, we will consider relations~(\ref{eq.1})
and~(\ref{eq.2}) as relations 
determining the invertible map from the operator $\hat A$ onto 
function $f_{\hat A}({\bf x})$.
Multiplying both sides of equation~(\ref{eq.2}) by the operator 
$\hat U({\bf x}')$ and taking trace, one has the consistency 
condition satisfied for the operators $\hat U({\bf x}')$ and
$\hat D({\bf x})$
\begin{equation}\label{eq.2'}
\mbox{Tr}\left[\hat U({\bf x}')\hat D({\bf x})\right]
=\delta\left({\bf x}'-{\bf x}\right).
\end{equation}
The consistency condition~(\ref{eq.2'}) follows from the relation
\begin{equation}\label{eq.2aa}
f_{\hat A}({\bf x})=\int K({\bf x}, {\bf x}')f_{\hat A}({\bf x}')
\,d{\bf x}'.
\end{equation}
The kernel in~(\ref{eq.2aa}) is equal to the standard Dirac 
delta-function if the set of functions
$f_{\hat A}({\bf x})$ is a complete set.
This is not the case for the tomographic map where the symbol of the 
operator is a homogeneous function of three variables.
In the case $\hat U(0)=\hat {\bf 1}$, symbol of the operator 
at ${\bf x}=0$, i.e. $f_{\hat A}(0)$ is equal to the trace of 
the operator $\hat A$, $~f_{\hat A}(0)=\mbox{Tr}\,(\hat A)$,
therefore we should require that our operators are trace-class,
in what follows we will not make this kind of qualifications 
any more.
There is some ambiguity in defining the operators
$\hat U({\bf x}')$ and $\hat D({\bf x})$.
One can make a scaling transform of the variables
${\bf x}$, which provides the corresponding scaling 
factor for redefining the operator $\hat D({\bf x})$.
If one defines the map for which the symbol of identity
operator $\hat {\bf 1}$ is equal to the unit function, 
the operator $\hat U({\bf x})$ 
satisfies the condition
\begin{equation}\label{eq.2'''}
\mbox{Tr}\,\hat U({\bf x})=1
\end{equation}
and 
the operator $\hat D({\bf x})$ 
satisfies the condition
\begin{equation}\label{eq.2''}
\int\hat D({\bf x})\,d{\bf x}=\hat {\bf 1}.
\end{equation}
In fact, we could consider relations of the form 
\begin{equation}\label{eq.3} 
\hat A\rightarrow f_{\hat A}({\bf x}) 
\end{equation}
and 
\begin{equation}\label{eq.4}
f_{\hat A}({\bf x})\rightarrow\hat A
\end{equation}
with the properties to be described below as
defining the map. The most important property is the
existence of associative product (star-product) of functions.
Some general considerations on star-products of functions are
made in appendices.
The operation of taking the trace in~(\ref{eq.1}) and
integrating in~(\ref{eq.2}) makes forms~(\ref{eq.3}) and  
(\ref{eq.4}) more concrete and gives the
possibility to describe properties
of the map. Let us discuss these properties. We introduce
the product (star-product) of two
functions $f_{\hat A}({\bf x})$ and $f_{\hat B}({\bf x})$
corresponding to two operators $\hat A$ and
$\hat B$ by the relations
\begin{equation}\label{eq.5}
f_{\hat A\hat B}({\bf x})=f_{\hat A}({\bf x})*
f_{\hat B} ({\bf x}):=\mbox{Tr}\left[\hat A\hat B\hat U({\bf x})
\right].
\end{equation}
Since the standard product of operators
on a Hilbert space is an associative product, i.e.
$\hat A(\hat B \hat C)=(\hat A\hat B)\hat C$, 
it is obvious that formula~(\ref{eq.5})
defines an associative product for the functions 
$f_{\hat A}({\bf x})$, i.e.
\begin{equation}\label{eq.6}
f_{\hat A}({\bf x})*\Big(f_{\hat B}({\bf x})
*f_{\hat C}({\bf x})\Big)=
\Big(f_{\hat A}({\bf x})*f_{\hat B}({\bf x})\Big)
*f_{\hat C}({\bf x}).
\end{equation}

\section{Ehrenfest picture as an example of a star-product
realization}

In the Ehrenfest picture of quantum evolution, we consider
quadratic functions on the Hilbert space of states defined by
$f_{\hat A}(\psi)=\langle\psi\mid\hat A\psi\rangle$.
Equations of motion can be written as 
\begin{equation}\label{Bep1}
i\hbar\,\dot f_{\hat A}=
\left\{f_{\hat H},f_{\hat A}\right\}
\end{equation}
with $\left\{f_{\hat B},f_{\hat A}\right\}(\psi)
:=f_{[\hat B,\hat A]}(\psi),
~\hbar$ being Planck constant and $\hat H$,
Hamiltonian operator. If we enlarge the picture to functions
on ${\cal H}$$\times$${\cal H}$, i.e. 
$f_{\hat A}(\psi_1,\psi_2)=\langle\psi_1\mid\hat A\psi_2\rangle$,
we can still write equations of motion in terms of Poisson brackets
of these functions on ${\cal H}$$\times$${\cal H}$.
Clearly the usual product (point-wise) of these quadratic functions 
is providing us with a `quartic function'.
It is therefore interesting to have, also for reasons of interpretation, 
a new product associating a quadratic function out of two quadratic
ones. Indeed, by setting 
$$
\hat{\bf 1}=\sum_n\mid\varphi_n\rangle\langle\varphi_n\mid\qquad
\mbox{or}\qquad 
\hat{\bf 1}=\int\mid\varphi_x\rangle\langle\varphi_x\mid\,dx
$$
we define
\begin{equation}\label{Bep2}
 f_{\hat A}(\psi_1,\psi_2)*f_{\hat B}(\psi_1,\psi_2):=
\sum_n \langle\psi_1\mid\hat A\mid \varphi_n\rangle
\langle\varphi_n\mid\hat B\psi_2\rangle
\end{equation}
or
\begin{equation}\label{Bep3}
 f_{\hat A}(\psi_1,\psi_2)*f_{\hat B}(\psi_1,\psi_2):=
\int dx\, \langle\psi_1\mid\hat A\mid\varphi_x\rangle
\langle\varphi_x\mid\hat B\psi_2\rangle
\end{equation}
as the case may be.

This new defined product is not `point-wise' and, nevertheless,
gives
\begin{equation}\label{Bep4}
i\hbar\,\dot f_{\hat A}=f_{\hat H}*f_{\hat A}
-f_{\hat A}*f_{\hat H}
\end{equation}
i.e. the same equations of motion can be described either in terms 
of the standard Poisson brackets or in terms of the commutator 
product associated with the star-product we have introduced.

As a matter of fact, if we use a numerable basis for ${\cal H}$,
operators are described by matrices with matrix elements $A_{ik}$ 
determined, if basis vectors in the Hilbert space are denoted by
$\mid i\rangle$ as 
\begin{equation}\label{eq.7}
A_{ik}=\langle i \mid \hat A \mid k\rangle.
\end{equation}
The standard rule of the matrix multiplication 
\begin{equation}\label{eq.8}
(AB)_{ij}=\sum_k A_{ik}B_{kj} 
\end{equation}
reproduces a star-product
for the functions of two discrete variables
\begin{equation}\label{eq.9}
A(i,k)\sim A_{ik} \qquad B(i,k)\sim B_{ik}.
\end{equation} 
The function $C(i,j)$ is therefore the star-product
of functions $A(i,j)$ and $B(i,j)$, i.e.
\begin{equation}\label{eq.10} 
C(i,j)=A(i,j)*B(i,j,) 
\end{equation}
if
\begin{equation}\label{eq.11}
C(i,j)=\sum_k A(i,k)B(k,j).
\end{equation}
Formulae~(\ref{eq.10}) and (\ref{eq.11}) provide
the composition rule for two functions
$A(i,j)$ and $B(i,j)$. The sum in 
equations~(\ref{eq.8}) and (\ref{eq.11}) can be considered 
as the integral, if the indices
$i,k$ are continuous ones. Thus the product of matrices
provides the simplest example of an associative
star-product of the matrix elements of matrices
considered as functions of position, e.g. 
$\langle x\mid A\mid x'\rangle$.
 This means that the introduction by Heisenberg of quantum
mechanics as the matrix mechanics in the early
days of quantum theory can be considered as a prototype of
star-product. This case is realized in formula~(\ref{eq.1}) 
by using the two-dimensional vector 
${\bf x}=(x_1,x_2)\in {\cal Z}^+$$\times$${\cal Z}^+$, where
$x_1=i$, $x_2=k$,  with $i$ and $k$ determining the row 
and column,
respectively, and the set of operators $\hat U({\bf x})$
is taken as
\begin{equation}\label{eq.12}
\hat U({\bf x})\sim\hat U(i,k)=\mid i\rangle\langle k\mid.
\end{equation}
The inverse formula~(\ref{eq.2}) can be considered,
if one uses the operator $\hat D({\bf x})\sim \hat D(i,k)$
in the same form 
\begin{equation}\label{eq.13} 
\hat D^\dagger({\bf x})=\mid i\rangle\langle k\mid
\end{equation}
i.e., in the simplest case, 
\begin{equation}\label{eq.14}
\hat D({\bf x})=\hat U^\dagger({\bf x})
\end{equation}
which means that the basic operators, defining
the functions in terms of operators, and the basic operators, defining
the operators in terms of the functions, coincide. 

We have restricted these considerations to
discrete basis on Hilbert space of states. When matrices of
observables in the position and momentum representations are needed,
they can be
also presented within the same framework by a simple replacement in 
(\ref{eq.7})--(\ref{eq.13}) of discrete indices with continuous
indices, i.e.$~i,k\rightarrow x,x'$ or $i,k\rightarrow p,p'$.

\section{Commutation relation and Heisenberg equation of motion} 

In view of (\ref{eq.1}), as we have already noticed,
the commutation bracket of two operators 
\begin{equation}\label{eq.15} 
\hat C=[\hat A, \hat B]=\hat A\hat B-\hat B\hat A 
\end{equation}
is mapped onto the Poisson bracket $f_{\hat C}({\bf x})$ of two
functions $f_{\hat A}({\bf x})$ and $f_{\hat B}({\bf x})$ by
means of the formula
\begin{equation}\label{eq.16} 
f_{\hat C}({\bf x})=\Big\{f_{\hat A}({\bf x}),
f_{\hat B}({\bf x})\Big\}_*=
\mbox{Tr}\left[[\hat A,\hat B]\hat U({\bf x})\right].
\end{equation}
Since the Jacobi identity is fulfilled for the
commutator of the operators, i.e.
\begin{equation}\label{eq.17}
\left[[\hat A, \hat B],\hat C\right]+
\left[[\hat B,\hat C],\hat A\right]
+\left[[\hat C,\hat A],\hat B\right]=0
\end{equation}
the Jacobi identity is also fulfilled for the
Poisson bracket of the functions $f_{\hat A}({\bf x})$ and
$f_{\hat B}({\bf x})$ defined by equation~(\ref{eq.15}). 

Since for the operators one has the derivation property
$$ \left[\hat A,\hat B\hat C\right]=
\left[\hat A,\hat B\right]\hat C+\hat B
\left[\hat A,\hat C\right]$$
the Poisson brackets~(\ref{eq.16}) reproduce this property
$$\fl 
\Big\{f_{\hat A}({\bf x}),f_{\hat B}({\bf x})*
f_{\hat C}({\bf x})\Big\}_*=\Big\{f_{\hat A}({\bf x}),
f_{\hat B}({\bf x})\Big\}_**f_{\hat C}({\bf x})+
f_{\hat B}({\bf x})*\Big\{f_{\hat A}({\bf x}),
f_{\hat C}({\bf x})\Big\}_*
$$
which qualifies it as a `quantum Poisson bracket' according to
Dirac~\cite{Dirac}.

In quantum mechanics, the evolution of observables
$\hat A$ can be described by Heisenberg equation of 
motion  
\begin{equation}\label{eq.18}
\dot{\hat A}=i[\hat H,\hat A]\qquad (\hbar=1)
\end{equation}
where $\hat H$ is the Hamiltonian of the system.
This equation can be rewritten in terms of the functions
$f_{\hat A}({\bf x})$ and $f_{\hat H}({\bf x})$, where 
\begin{equation}\label{eq.19}
f_{\hat H}({\bf x})=\mbox{Tr}\left[\hat H\hat U({\bf x})\right]
\end{equation}
corresponds to the Hamiltonian, in the form 
\begin{equation}\label{eq.20}
\dot f_{\hat A}({\bf x},t)=i\Big\{f_{\hat H}({\bf x},t),
f_{\hat A}({\bf x},t)\Big\}_*
\end{equation}
with the Poisson bracket defined by
equation~(\ref{eq.16}) using the star-product given by
equation~(\ref{eq.5}).

\section{Relation between different maps}

Let us suppose that there exist other maps analogous
to the ones given by equations~(\ref{eq.1}) and~(\ref{eq.2}).
Let us choose two different ones.
One map is described by a vector ${\bf x}=(x_1,x_2,\ldots,x_n)$, 
operator $\hat U({\bf x})$
and operator $\hat D({\bf x})$ in
formulae~(\ref{eq.1}) and (\ref{eq.2}).
Another map is described by a vector
${\bf y}=(y_1,y_2,\ldots,y_m)$ and operators
$\hat U_1({\bf x})$ and
$\hat D_1({\bf y})$ in (\ref{eq.1}) and (\ref{eq.2}), 
respectively, i.e., for given operator $\hat A$, one has 
the function
\begin{equation}\label{eq.21} 
\phi_{\hat A}({\bf y})=\mbox{Tr}\left[\hat A\hat U_1({\bf y})\right]
\end{equation}
and the inverse relation
\begin{equation}\label{eq.22}
\hat A=\int\phi_A({\bf y})\hat D_1({\bf y})~d{\bf y}. 
\end{equation}
One can obtain a relation of the
function $f_{\hat A}({\bf x})$ with the function 
$\phi_{\hat A}({\bf y})$ in the form
\begin{equation}\label{eq.23}
\phi_{\hat A}({\bf y})=\int f_{\hat A}({\bf x})\,\mbox{Tr}\left[
\hat D({\bf x})\hat U_1({\bf y})\right]\,d{ \bf x}
\end{equation}
and the inverse relation 
\begin{equation}\label{eq.24}
f_{\hat A}({\bf x})=
\int\phi_{\hat A}({\bf y})\,\mbox{Tr}\left[\hat D_1({\bf y})
\hat U({\bf x})\right]\,d{\bf y}.
\end{equation}
We see that functions $f_{\hat A}({\bf x})$ and
$\phi_{\hat A}({\bf y})$ corresponding to different
maps are connected
by means of the invertible integral transform given
by equations~(\ref{eq.23}) and (\ref{eq.24}).
These transforms are determined by means of intertwining 
kernels in (\ref{eq.23}) and (\ref{eq.24})
\begin{equation}\label{eq.28aa}
K_1({\bf x},{\bf y})=\mbox{Tr}\,\Big[\hat D({\bf x})
\hat U_1({\bf y})\Big ]
\end{equation}
and 
\begin{equation}\label{eq.28bb}
K_2({\bf x},{\bf y})=\mbox{Tr}\,\Big[\hat D_1({\bf y})
\hat U({\bf x})\Big].
\end{equation}

\section{Star-product as a composition rule for two symbols} 

Using formulae~(\ref{eq.1}) and (\ref{eq.2}), one can write
down a composition rule for two symbols $f_{\hat A}({\bf x})$
and $f_{\hat B}({\bf x})$, which determines the star-product
of these symbols. The composition rule is described by the
formula 
\begin{equation}\label{eq.25}
f_{\hat A}({\bf x})*f_{\hat B}({\bf x})=
\int f_{\hat A}({\bf x}'')f_{\hat B}({\bf x}')
K({\bf x}'',{\bf x}',{\bf x})\,d{\bf x}'\,d{\bf x}''.
\end{equation} 
The kernel in the integral of (\ref{eq.25}) is
determined by the trace of product of the basic operators,
which we use to construct the map 
\begin{equation}\label{eq.26} 
K({\bf x}'',{\bf x}',{\bf x})=
\mbox{Tr}\left[\hat D({\bf x}'')\hat D({\bf x}')
\hat U({\bf x})\right].
\end{equation} 
In the following sections, we calculate this kernel 
for some important examples of the map. 

Formula~(\ref{eq.26}) can be extended to the case of 
the star-product of $N$ symbols of operators
$\hat A_1,\hat A_2,\ldots,\hat A_N$.
Thus one has
\begin{eqnarray}\label{eq.26'}
\fl W_{\hat A_1}({\bf x})*W_{\hat A_2}({\bf x})*\cdots *
W_{\hat A_N}({\bf x})=\int W_{\hat A_1}({\bf x}_1)
W_{\hat A_2}({\bf x}_2)\cdots W_{\hat A_N}({\bf x}_N)\nonumber\\
\fl \times
K\left({\bf x}_1,{\bf x}_2,\ldots,{\bf x}_N,{\bf x}\right)
\,d{\bf x}_1\,d{\bf x}_2\cdots \,d{\bf x}_N
\end{eqnarray}
where the kernel has the form
\begin{equation}\label{eq.26''}
K\left({\bf x}_1,{\bf x}_2,\ldots,{\bf x}_N,{\bf x}\right)=
\mbox{Tr}\left[\hat D({\bf x}_1)\hat D({\bf x}_2)
\cdots\hat D({\bf x}_N)\hat U({\bf x})\right].
\end{equation}
Since this kernel determines the associative star-product
of $N$ symbols, it can be expressed in terms of the kernel
of the star-product of two symbols.
The trace of an operator $\hat A^{N}$ is determined by 
the kernel as follows
\begin{eqnarray}\label{eq.26'''}
&&\mbox{Tr}\, \hat A^N=\int W_{\hat A}({\bf x}_1)
W_{\hat A}({\bf x}_2)\cdots W_{\hat A}({\bf x}_N)
\nonumber\\
&&\times \mbox{Tr}\left[\hat D({\bf x}_1)\hat D({\bf x}_2)
\cdots\hat D({\bf x}_N)\right]
\,d{\bf x}_1\,d{\bf x}_2\cdots \,d{\bf x}_N.
\end{eqnarray}
When the operator $\hat A$ is a density operator of a quantum state,
formula~(\ref{eq.26'''}) determines the generalized purity parameter 
of the state. 
When the operator $\hat A$ is equal to product of two density operators
and $N=1$, formula (\ref{eq.26'''}) determines the fidelity.
 
\section{Weyl symbol}

In this section, we will consider a known 
example of the Heisenberg--Weyl-group representation. 
As operator $\hat U({\bf x})$, we take the Fourier
transform of displacement operator
$\hat D(\xi)$
\begin{equation}\label{eq.27}
\hat U({\bf x})=\int
\exp\left(\frac{x_1+ix_2}{\sqrt 2}
\mbox{\boldmath$\xi$}^*-\frac{x_1-i x_2}{\sqrt 2}
\mbox{\boldmath$\xi$}\right)
\hat D(\mbox{\boldmath$\xi$})\pi^{-1}~d^2
\mbox{\boldmath$\xi$}
\end{equation}
where $\mbox{\boldmath$\xi$}$ is a complex number,
$\mbox{\boldmath$\xi$}=\xi_1+i\xi_2$, 
and the vector ${\bf x}=(x_1,x_2)$
can be considered as ${\bf x}=(q,p)$, with $q$ and $p$ being 
position and momentum. 
One can see that $\mbox{Tr}\,\hat U({\bf x})=1$.
The displacement operator may be expressed through creation and
annihilation operators in the form 
\begin{equation}\label{eq.28} 
\hat D(\mbox{\boldmath$\xi$})=\exp(\mbox{\boldmath$\xi$}
\hat a^\dagger-\mbox{\boldmath$\xi$}^*\hat a).
\end{equation}
The displacement operator is used to create coherent
states from the vacuum state.
For creation and annihilation operators, one has 
\begin{equation}\label{eq.29}
\hat a=\frac{\hat q+i\hat p}{\sqrt 2} \qquad
\hat a^\dagger=\frac{\hat q- i\hat p}{\sqrt 2}
\end{equation}
where $\hat q$ and $\hat p$ may be thought as  
coordinate and momentum  operators
for the carrier space of an harmonic 
oscillator.
The operator $\hat a$ and its Hermitian conjugate 
$\hat a^\dagger$ satisfy the boson commutation relation 
$[\hat a,\hat a^\dagger]=\hat {\bf 1}.$

Let us introduce the Weyl symbol for an arbitrary
operator $\hat A$ using the definition~(\ref{eq.1})
\begin{equation}\label{eq.30}
W_{\hat A}({\bf x})=\mbox{Tr}\left[\hat A\hat U({\bf x})
\right]
\end{equation}
the form of operator $\hat U({\bf x})$ is given by
equation~(\ref{eq.27}). One can check that Weyl symbols
of the identity operator $\hat {\bf 1}$, 
position operator $\hat q$
and momentum operator $\hat p$ have the form
\begin{equation}\label{eq.30'}
W_{\hat {\bf 1}}(q,p)=1\qquad
W_{\hat q}(q,p)=q\qquad
W_{\hat p}(q,p)=p.
\end{equation}
The inverse transform, which expresses the operator $\hat A$
through its Weyl symbol, is of the form
\begin{equation}\label{eq.31}
\hat A=\int
W_{\hat A}({\bf x})\hat U({\bf x})\,\frac{d{\bf x}}{2\pi}\,.
\end{equation}
One can check that for $W_{\hat {\bf 1}}({\bf x})=1$,
formula~(\ref{eq.31}) reproduces the identity operator, i.e.
\begin{equation}\label{eq.31'}
\int \hat U({\bf x})\,\frac{d{\bf x}}{2\pi}=\hat {\bf 1}.
\end{equation}
Comparing (\ref{eq.31}) with~(\ref{eq.2}), one can see that 
the operator $\hat D({\bf x})$ in formula~(\ref{eq.2}) is
connected with $\hat U({\bf x})$ by the relation
\begin{equation}\label{eq.32}
\hat D({\bf x})=\frac{\hat U({\bf x})}{2\pi}\,.
\end{equation}
Let us consider now the star-product of two
Weyl symbols (it is usually called Moyal star-product). 
If one takes two operators $\hat A_1$ and
$\hat A_2$, which are expressed through Weyl
symbols by formulae
\begin{equation}\label{eq.33}                                         
\fl \hat A_1=\int
W_{\hat A_1}({\bf x}')\hat U({\bf x}')\,\frac{d{\bf x}'}{2\pi}
\qquad
\hat A_2=\int W_{\hat A_2}({\bf x}'')
\hat U({\bf x}'')\,\frac{d{\bf x}''}{2\pi}
\end{equation}
with vectors ${\bf x}'=(x_1',x_2')$ and ${\bf x}''=(x_1'',x_2'')$,
the operator $\hat A$ (product of operators
$\hat A_1$ and $\hat A_2)$ has Weyl symbol given by
\begin{eqnarray}
\fl W_{\hat A}({\bf x})=\mbox{Tr}\,
\Big[\hat A\hat U({\bf x})\Big]=\frac{1}{4\pi^5}\,\int
d{\bf x}'\,d{\bf x}''\,d^2{\bf\xi}\,d^2{\bf\xi}'\,d^2{\bf\xi}''\,
W_{\hat A_1}({\bf x}')W_{\hat A_2}({\bf x}'')\nonumber\\
\fl \times\exp\Big\{2^{-1/2}\Big[(\xi_1'-i\xi_2')(x_1'+ix_2')-
(\xi_1'+i\xi_2')(x_1'-ix_2')\nonumber\\
\fl +(\xi_1''-i\xi_2'')(x_1''+ix_2'')
-(\xi_1''+i\xi_2'')(x_1''-ix_2'')+
(\xi_1-i\xi_2)(x_1+ix_2)\nonumber\\
\fl -(\xi_1+i\xi_2)(x_1-ix_2)\Big]\Big\}\,
\mbox{Tr}\left[\hat D(\mbox{\boldmath$\xi$}')
\hat D(\mbox{\boldmath$\xi$}'')
\hat D(\mbox{\boldmath$\xi$})\right] \label{eq.34}
\end{eqnarray}
where $\mbox{\boldmath$\xi$}
=\xi_1+i\xi_2$, with $~\mbox{\boldmath$\xi$}'=\xi_1'+i\xi_2'$
and $~\mbox{\boldmath$\xi$}''=\xi_1''+i\xi_2''$.
Using properties of displacement operators  
\begin{equation}\label{eq.35}
\fl \hat D(\mbox{\boldmath$\xi$}')\hat D
(\mbox{\boldmath$\xi$}'')=
\hat D(\mbox{\boldmath$\xi$}'+\mbox{\boldmath$\xi$}'')
\exp\Big(i\mbox{ Im}\,(\mbox{\boldmath$\xi$}'
\mbox{\boldmath$\xi$}''^*)\Big) \qquad 
\mbox{Tr}\left[\hat D(\mbox{\boldmath$\xi$})\right]
=\pi\delta^2(\mbox{\boldmath$\xi$}) 
\end{equation}
one has for the star-product of two Weyl symbols the
following formula 
\begin{eqnarray}
&&W_{\hat A}({\bf x})=W_{\hat A_1}({\bf x})*
W_{\hat A_2}({\bf x})
=\int \frac{d{\bf x}'\,d{\bf x}''}{\pi^2}\,
W_{\hat A_1}({\bf x}')
W_{\hat A_2}({\bf x}'')\nonumber\\
&&\times\exp\Big\{2i\Big[(x_2'-x_2)(x_1-x_1'')
+(x_1'-x_1)(x_2''-x_2)\Big]\Big\}.
\label{eq.36}
\end{eqnarray}
This formula coincides with~(\ref{eq.25}), in which one
uses the kernel 
\begin{equation}\label{eq.37}
\fl K\left({\bf x}'',{\bf x}',{\bf x}\right)=
\pi^{-2}\exp\Big\{2i\Big[(x_2'-x_2)(x_1-x_1'')+
(x_1'-x_1)(x_2''-x_2)\Big]\Big\}.
\end{equation}
If we consider $\hat A$ to be the density operator
$\hat\rho$, the corresponding Wigner function
$W({\bf x})$ is
\begin{equation}\label{eq.38}
W_{\hat\rho}({\bf x})\equiv W({\bf x})
=\mbox{Tr}\,\Big[\hat\rho\hat U({\bf x})\Big]
\end{equation}
the inverse transform reads
\begin{equation}\label{eq.39}
\hat\rho=\int
W_{\hat\rho}({\bf x})\hat U({\bf x})\,\frac{d{\bf x}}{2\pi}\,.
\end{equation} 
The star-product of two Wigner functions
$ W_{\hat\rho_1}({\bf x})*W_{\hat\rho_2}({\bf x})$, which 
corresponds to the operator $\hat\rho$ (the product of density
operators $\hat\rho_1$ and $\hat\rho_2$) is the
function $W({\bf x})$ determined as 
\begin{equation}\label{eq.40} 
\fl W({\bf x})=\mbox{Tr}\,\Big[\hat\rho_1\hat\rho_2
\hat U({\bf x})\Big]
=\int d{\bf x}'\,d{\bf x}''\,
W_{\hat \rho_1}({\bf x}')W_{\hat \rho_2}({\bf x}'')
K\left({\bf x}',{\bf x}'',{\bf x}\right)
\end{equation}
where $K({\bf x}',{\bf x}'',{\bf x})$ is given
by~(\ref{eq.37}). 

\section{Star-product of $s$-ordered symbols}

Following~\cite{CahillGlauber67,MarmoSemicOpt},
let us define the $s$-ordered symbol  
function $W_{\hat A}({\bf x},s)$, which 
corresponds to some operator $\hat A$ in
the general case of Heisenberg--Weyl group 
\begin{equation}\label{s1} 
W_{\hat A}({\bf x},s)=
\mbox{Tr}\left[\hat A\hat U({\bf x},s)\right]
\end{equation}
with a real parameter $s$, real vector ${\bf x}=(x_1,x_2)$
and the operator $\hat U({\bf x},s)$ of the form
\begin{equation}\label{s2} 
\hat U({\bf x},s)=\frac{2}{1-s}\,\hat D(\alpha_{\bf x})
\,q^{\hat a^\dagger\hat a}(s)\,\hat D(-\alpha_{\bf x})
\end{equation}
now the displacement operator reads
\begin{equation}\label{s3} 
\hat D(\alpha_{\bf x})=\exp\left(\alpha_{\bf x}
\hat a^\dagger-\alpha^*_{\bf x}\hat a\right).
\end{equation}
We have also
\begin{equation}\label{s4} 
\fl \alpha_{\bf x}=x_1+ix_2\qquad
\alpha^*_{\bf x}=x_1-ix_2\qquad
x_1=\frac{q}{\sqrt 2}\qquad
x_2=\frac{p}{\sqrt 2}
\end{equation}
while the parameter $q(s)$ is
\begin{equation}\label{s5}
q(s)=\frac{s+1}{s-1}\,.
\end{equation}
Thus we rescaled variables $x_1$ and $x_2$ in comparison with
equation~(\ref{eq.27}).
The coefficient in equation~(\ref{s2}) provides
the property 
$\mbox{Tr}\left[\hat U({\bf x},s)\right]=1$,
which means that the symbol of the identity operator equals 1.

One can see that
\begin{equation}\label{s6}
q(-s)=q^{-1}(s)
\end{equation}
and, in view of the commutation relation of creation 
and annihilation operators 
$\hat a\hat a^{\dagger}-\hat a^{\dagger}\hat a=1$,
one has the following relation:
\begin{eqnarray}\label{s7}
\exp\left(\alpha_{\bf x_1}
\hat a^\dagger-\alpha^*_{\bf x_1}\hat a\right)
\exp\left(\alpha_{\bf x_2}
\hat a^\dagger-\alpha^*_{\bf x_2}\hat a\right)\nonumber\\
\fl =\exp\left[\left(\alpha_{\bf x_1}+\alpha_{\bf x_2}\right)
\hat a^{\dagger}-\left(\alpha^*_{\bf x_1}
+\alpha^*_{\bf x_2}\right)\hat a
+\frac{1}{2}\left(\alpha^*_{\bf x_2}\alpha_{\bf x_1}
-\alpha^*_{\bf x_1}\alpha_{\bf x_2}\right)\right].
\end{eqnarray}
The relation can be obtained using the 
Baker--Campbell--Hausdorf formula
$$e^Ae^B=e^{A+B+[A,B]/2},$$
in which operators $A$ and $B$ commute with the operator
$[A,B]$.
The operator $\hat U({\bf x},s)$ has the property 
\begin{equation}\label{s8} 
\hat U({\bf x},-s)=\frac{2}{1+s}\,\hat D(\alpha_{\bf x})\,
q^{-\hat a^\dagger\hat a}(s)\,\hat D(-\alpha_{\bf x}).
\end{equation}
One checks that
\begin{equation}\label{s9} 
\mbox{Tr}\left[\hat U({\bf x_1},-s)
\hat U({\bf x_2},s)\right]=\pi\,\frac{1-s}{1+s}\,
\delta\left({\bf x}_1-{\bf x}_2\right).
\end{equation}
Due to relation~(\ref{s9}), the expression for the
operator $\hat A$ is given by the relation inverse to (\ref{s1}) 
\begin{equation}\label{s10}
\hat A=\frac{1}{\pi}\,\frac{1+s}{1-s}\,\int W_{\hat A}({\bf x},s)
\hat U({\bf x},-s)\,d({\bf x}).
\end{equation}
This means that, for $s$-ordered symbols, the operator 
$\hat D({\bf x})$ in the general formula~(\ref{eq.2}) takes the form
\begin{equation}\label{s11}
\hat D({\bf x})\Longrightarrow\frac{1}{\pi}\,\frac{1+s}{1-s}
\,\hat U({\bf x},-s).
\end{equation}
For $s=0$, one has
\begin{eqnarray}
\hat U({\bf x})&=&2\hat D(\alpha_{\bf x})
(-1)^{\hat a^\dagger\hat a}
\hat D(-\alpha_{\bf x})\label{s12}\\
\hat D({\bf x})&=&\frac{2}{\pi}\,\hat D(\alpha_{\bf x})
(-1)^{\hat a^\dagger\hat a}
\hat D(-\alpha_{\bf x}).
\label{s13}
\end{eqnarray}
Due to the rescaling of the vector ${\bf x}$, instead of
equation~(\ref{eq.27}), one has 
$$\hat U({\bf x})=\pi\hat D({\bf x})$$
which is compatible with relation~(\ref{eq.32}) written for 
the vector ${\bf x}=(q,p)$.  
The operator $(-1)^{\hat a^\dagger\hat a}$
is the parity operator $(-1)^{\hat a^\dagger\hat a}=\hat P$,
with the matrix elements given in the position
representation by the formula
\begin{equation}\label{s13'}
\langle x\mid\hat P\mid y\rangle=\delta(x+y).
\end{equation}
One can check that the following relation holds true
\begin{equation}\label{s13''}
\hat P\exp\left(\alpha_{\bf x}\hat a^\dagger
-\alpha^*_{\bf x}\hat a\right)=
\exp\left(\alpha^*_{\bf x}\hat a
-\alpha_{\bf x}\hat a^\dagger\right)\hat P.
\end{equation}
Since $\hat P\hat P=\hat{\bf 1}$, one arrives at
\begin{equation}\label{s14}
\hat U({\bf x})\hat D({\bf x}')=\frac{4}{\pi}\,
\hat D(2\alpha_{\bf x})
\hat D(-2\alpha'_{\bf x})
\end{equation}
and due to formula~(\ref{eq.35}) one obtains
\begin{equation}\label{s15} 
\mbox{Tr}\left[\hat U({\bf x})
\hat D({\bf x}')\right]=
\delta\left({\bf x}-{\bf x}'\right).
\end{equation}
If one uses the coordinates of vector ${\bf x}$ given by 
equation~(\ref{s4}), the symbols of the position operator 
$\hat q$ and the momentum operator $\hat p$ will be equal 
to $q$ and $p$, respectively, for arbitrary parameter $s$. 

For $s=0$, 
formula~(\ref{s1}) provides Weyl symbol of the 
operator $\hat A$ considered in the previous section.
One can check this directly using the matrix elements 
of the operator~(\ref{s12}) in the position representation,
since the matrix elements of the displacement operator 
are given by the formula 
\begin{equation}\label{s16}
\fl \langle x\mid\hat D(\alpha_{\bf x})
\mid y\rangle=\exp\left[\frac{\alpha_{\bf x}
-\alpha^*_{\bf x}}{\sqrt 2}\,x-\frac{\alpha^2_{\bf x}
-\alpha^{*2}_{\bf x}}{4}\right]
\delta\left(x-y-\frac{\alpha_{\bf x}
+\alpha^*_{\bf x}}{\sqrt 2}\right)
\end{equation}
and the kernel of the operator~(\ref{s12}) reads
\begin{equation}\label{s17}
\langle x\mid\hat U({\bf x})\mid t\rangle=
2\int \langle x\mid\hat D(2\alpha_{\bf x})
\mid y\rangle\delta\left(y+t\right)\,dy.
\end{equation}
To calculate the kernel for the star-product of 
$s$-ordered symbols, one needs to calculate the 
trace of the product of two operators
\begin{equation}\label{s18}
Z= \mbox{Tr}\left[\hat D\left(\alpha,\tilde \alpha^*\right)
q^{\hat a^\dagger\hat a}\right]
\end{equation}
where $q$ is a real parameter and the operator 
$\hat D\left(\alpha,\tilde\alpha^*\right)$ (deformed displacement 
operator) has the same form as the unitary displacement 
operator creating the coherent state from vacuum
\begin{equation}\label{s19}
\hat D\left(\alpha,\tilde\alpha^*\right)
=\exp\left(\alpha \hat a^{\dagger}-\tilde\alpha^* \hat a\right)
\end{equation}
but now we consider the complex numbers $\alpha$ and
$\tilde\alpha^*$ as arbitrary and independent complex numbers.

A relation analogous to (\ref{s7}) is valid for the 
operators~(\ref{s19}). In view of the completeness relation
for coherent states
\begin{equation}\label{s20}
\frac{1}{\pi}\int d^2\beta\,\mid\beta\rangle
\langle\beta\mid=1 \qquad d^2\beta=d\beta_1\,d\beta_2
\end{equation}
using the action of the operator $q^{\hat a^\dagger\hat a}$
onto coherent states
\begin{equation}\label{s21}
q^{\hat a^\dagger\hat a}\mid\beta\rangle=\exp\left(
\frac{q^2-1}{2}|\beta|^2\right)\mid q\beta\rangle
\end{equation}
after calculating the Gaussian integral in 
equation~(\ref{s18}), one obtains
\begin{equation}\label{s22}
Z=\frac{1}{1-q}\,\exp\left[-\left(
\frac{q}{1-q}+\frac{1}{2}\right)\alpha\tilde\alpha^*\right].
\end{equation}

To calculate the kernel of the star-product, one needs other 
properties of the function of creation and annihilation
operators. One can use the following relations:
\begin{equation}\label{s23}
q^{\hat a^\dagger\hat a}\hat a=\hat a
q^{\hat a^\dagger\hat a-1}\qquad
q^{\hat a^\dagger\hat a}\hat a^\dagger=\hat a^\dagger
q^{\hat a^\dagger\hat a+1}
\end{equation}
which induce the following relations for arbitrary functions
of the creation and annihilation operators:
\begin{equation}\label{s24}
q^{\hat a^\dagger\hat a}f(\hat a)=f(q^{-1}\hat a)
q^{\hat a^\dagger\hat a}\qquad
q^{\hat a^\dagger\hat a}f(\hat a^\dagger)=
f(q\hat a^\dagger)q^{\hat a^\dagger\hat a}.
\end{equation}
For the deformed displacement operator~(\ref{s19}),
the following relation holds
\begin{equation}\label{s25}
q^{\hat a^\dagger\hat a}\hat D\left(\alpha,\tilde\alpha^*\right)
=\hat D\left(\alpha_q,\tilde\alpha^*_q\right)
q^{\hat a^\dagger\hat a}
\end{equation}
where
\begin{equation}\label{s26}
\alpha_q=q\alpha\qquad
\tilde\alpha^*_q=q^{-1}\tilde\alpha^*.
\end{equation}
In view of this notation, one can rewrite the operator~(\ref{s2})
using the following replacement:
$$
{\bf x}\rightarrow\alpha\qquad 
\hat U({\bf x},s)\rightarrow\hat U(\alpha,q)\qquad 
q(s)\rightarrow q
$$
where one has
\begin{equation}\label{s27}
\hat U(\alpha,q)=(1-q)\hat D(\alpha)
\,q^{\hat a^\dagger\hat a}\,\hat D(-\alpha).
\end{equation}
Introducing the operator
\begin{equation}\label{s28}
\hat D(\alpha,q)=\frac{1}{\pi}\left(1-q^{-1}\right)
\hat D(\alpha)\,q^{-\hat a^\dagger\hat a}\,\hat D(-\alpha)
\end{equation}
one has 
\begin{equation}\label{s29}
\widetilde Z= \mbox{Tr}\left[\hat U(\alpha,q)\hat D(\beta,q)\right]
=\delta^{(2)}(\alpha-\beta).
\end{equation}
In the following formula
\begin{eqnarray}\label{s30}
W_{\hat A}(\alpha_N)&=&\mbox{Tr}\left[
\hat A_1\hat A_2\cdots\hat A_{N-1}
\hat U(\alpha_N,q)\right]\nonumber\\
&=&\int K\left(\alpha_1,\alpha_2,\ldots,\alpha_N\right)
\left[\prod_{k=1}^{N-1}W_{\hat A_k}(\alpha_k,q)
\,d^2\alpha_k\right]
\end{eqnarray}
the kernel of the star-product of $(N-1)$ symbols has 
the form
\begin{equation}\label{s31}
K\left(\alpha_1,\alpha_2,\ldots,\alpha_N\right)=
\mbox{Tr}\left[\hat U(\alpha_N,q)
\prod_{k=1}^{N-1}\hat D(\alpha_k,q)\right].
\end{equation}
Since the kernel is a Gaussian function, it can be calculated
using its particular form only for two nonzero values of
$\alpha_j,\alpha_{j+n}$. Employing the method elaborated,
in view of formulae~(\ref{s24})--(\ref{s26}), one can calculate 
the kernel, which for star-product of two symbols reads
\begin{eqnarray}\label{s27'}
&&K\left(\alpha_1,\alpha_2,\alpha_3\right)=\frac{(1-q)(1-q^{-1})}{\pi^2}
\,\exp\Big[(q-q^{-1})|\alpha_3|^2
+(q-1)\alpha_1\alpha_2^*\nonumber\\
&&+(1-q^{-1})\alpha_2\alpha_1^*
+(q^{-1}-1)\alpha_2\alpha_3^*+(1-q)\alpha_3\alpha_2^*
\nonumber\\
&&+(q^{-1}-1)\alpha_3\alpha_1^*+(1-q)\alpha_1\alpha_3^*\Big].
\end{eqnarray}
The kernel $K\left(\alpha_1,\alpha_2,\ldots,\alpha_N\right)$
can be calculated explicitly using algebraic relations we 
elaborated and commutation relations we employed. The result 
of calculations follows
\begin{eqnarray}\label{s28'}
&&K\left(\alpha_1,\alpha_2,\ldots,\alpha_N\right)=
\frac{1-q}{1-\widetilde q}\,
\frac{(1-q^{-1})^{N-1}}{\pi^{N-1}}
\nonumber\\
&&\times\exp\left[-\sum_{i<j}^N
\mbox{\boldmath$\alpha$}_iM_{ij}
\mbox{\boldmath$\alpha$}_j-\sum_{i=1}^N
\mbox{\boldmath$\alpha$}_i
\Big(f(\widetilde q)\sigma_x-\hat d_i\Big) 
\mbox{\boldmath$\alpha$}_i\right]
\end{eqnarray}
where the 2-vector $\mbox{\boldmath$\alpha$}_i$, parameter
$\widetilde q$ and function $f(\widetilde q)$ are
$$
\mbox{\boldmath$\alpha$}_i=\left(\begin{array}{c}
\alpha_i\\
\alpha_i^*\end{array}\right)\qquad 
\widetilde q=q^{2-N}\qquad q=\frac{s+1}{s-1}\qquad
f(\widetilde q)=\frac{1}{2}
\left(\frac{\widetilde q}{1-\widetilde q}
+\frac{1}{2}\right).
$$
The matrices $M_{ij}$ have the form
\begin{eqnarray*}
&&M_{ij}=\frac{(q-1)(q^{-1}-1)}{1-q^{2-N}}\left(
\begin{array}{clcr}
0&q^{2-N+j-i}\\
q^{i-j}&0\end{array}\right)\qquad j<N
\nonumber\\
&&M_{iN}=-M_{iN-1}\qquad i< N.
\end{eqnarray*}
The antidiagonal matrices 
$\hat d_1=\hat d_2=\cdots=\hat d_{N-1}$
and $\hat d_N$ are such 
that the kernel~(\ref{s28}) can be rewritten in
terms of the complex numbers 
$\alpha_i \,\left(i=1,2,\ldots,N\right)$ as
\begin{eqnarray}\label{s29'}
&&K\left(\alpha_1,\alpha_2,\ldots,\alpha_N\right)=
\frac{1-q}{1-q^{2-N}}\,
\frac{(1-q^{-1})^{N-1}}{\pi^{N-1}}
\nonumber\\
&&\times\exp\left\{\sum_{j>i}^{N-1}\,\sum_{i=1}^{N-1}
\frac{(q-1)(1-q^{-1})}{1-q^{2-N}}
\left(q^{j-i+2-N}\alpha_i\alpha_j^*+
q^{i-j}\alpha_j\alpha_i^*\right)\right.\nonumber\\
&&\left.+\sum_{i=1}^{N-1}
\frac{(1-q)(1-q^{-1})}{1-q^{2-N}}
\left(q^{1-i}\alpha_i\alpha_N^*+
q^{i+1-N}\alpha_N\alpha_i^*\right)\right.\nonumber\\
&&\left.+\frac{|\alpha_i|^2}{2}
\left[q^{-1}-q-\frac{q^{2-N}+1}{1-q^{2-N}}
\,(1-q)(1-q^{-1})\right]\right.\nonumber\\
&&\left. +\frac{|\alpha_N|^2}{2}
\left[q-q^{-1}-\frac{q^{2-N}+1}{1-q^{2-N}}
\,(1-q)(1-q^{-1})\right]\right\}.
\end{eqnarray}
Thus, we got a Gaussian form for the kernel of the star-product 
of $(N-1)$ operators. In the case $q=-1$, the kernel provides
the expression for the star-product of $(N-1)$ Weyl symbols. 
For $N=3$, the kernel reproduces equation~(\ref{s27'}).

When the operator $\hat A$ is a density operator 
$\hat\rho$, the purity parameter $\mu_0$ of the quantum state 
is defined in terms of the symbol of the operator $\hat\rho^2$
by the formula
\begin{equation}\label{s30'}
\mu_0=\int W_{\hat\rho}(\alpha_1)W_{\hat\rho}(\alpha_2)
\,\mbox{Tr}\,\Big[\hat D(\alpha_1,q),\hat D(\alpha_2,q)\Big]
\,d^2\alpha_1\,d^2\alpha_2.
\end{equation}
The other purity parameters 
$\mu_{N-2}=\mbox{Tr}\,\hat\rho^N$
are given by the formula 
\begin{eqnarray}\label{s31'}
&&\mu_{N-2}=\int W_{\hat\rho}(\alpha_1)W_{\hat\rho}(\alpha_2)
\cdots W_{\hat\rho}(\alpha_N)\nonumber\\
&&\times\mbox{Tr}\,\Big[\hat D(\alpha_1,q),\hat D(\alpha_2,q),
\ldots,\hat D(\alpha _N,q)\Big]
\,d^2\alpha_1\,d^2\alpha_2\cdots d^2\alpha_N
\end{eqnarray}
where the purity kernel reads
\begin{equation}\label{s31''}
\fl\mbox{Tr}\,\Big[\hat D(\alpha_1,q)\hat D(\alpha_2,q)\cdots
\hat D(\alpha_N,q)\Big]=
\frac{\pi}{(1-q)(1-q^{-1})}\,
K\left(\alpha_1,\alpha_2,\ldots,\alpha_N,0,0\right)
\end{equation} 
where the function $K$ is given by (\ref{s29'}) with
$\alpha_{N+1}=\alpha_{N+2}=0$.

For $s=0$ (Weyl representation), the kernel was 
calculated in~\cite{SudarJRLR}.

\section{Tomographic representation}

In this section, we will consider an example
of the probability representation of quantum mechanics. In
the probability representation of quantum mechanics, the
state is described by a family of probabilities.
According to the general scheme one can introduce for the
operator $\hat A$ the function $f_{\hat A}({\bf x})$, where
${\bf x}=(x_1,x_2,x_3)\equiv (X,\mu,\nu)$,
which we denote here as
$w_{\hat A}(X,\mu,\nu)$ depending on the position $X$ and
the reference frame parameters $\mu$ and $\nu$
\begin{equation}\label{eq.53}
w_{\hat A}(X,\mu,\nu)=\mbox{Tr}\left[\hat A
\hat U({\bf x})\right].
\end{equation}
We call the function $w_{\hat A}(X,\mu,\nu)$ the tomographic 
symbol of the operator $\hat A$. The operator $\hat U(x)$ is 
given by
\begin{eqnarray}\label{eq.54}
\fl 
\hat U({\bf x})\equiv \hat U(X,\mu,\nu)=
\exp\left(\frac{i\lambda}{2}\left(\hat q\hat p
+\hat p\hat q\right)\right)
\exp\left(\frac{i\theta}{2}\left(\hat q^2
+\hat p^2\right)\right)
\mid X\rangle\langle X\mid\nonumber\\
\qquad ~\times\exp\left(-\frac{i\theta}{2}\left(\hat q^2
+\hat p^2\right)\right)          
\exp\left(-\frac{i\lambda}{2}\left(\hat q\hat p
+\hat p\hat q\right)\right)\nonumber\\
\qquad =\hat U_{\mu\nu}\mid X\rangle\langle  
X\mid\hat U_{\mu\nu}^\dagger.
\end{eqnarray}
The angle $\theta$ and parameter $\lambda$ in terms of the
reference frame parameters are given by
$$
\mu=e^{\lambda}\cos\theta
\qquad \nu=e^{-\lambda}\sin\theta.
$$
Moreover, $\hat q$ and $\hat p$ are position and momentum 
operators,
\begin{equation}\label{eq.54'}
\hat q\mid X\rangle=X\mid X\rangle
\end{equation}
and
$\mid X\rangle\langle X\mid$ is the projection density. 
One has the canonical transform of quadratures
$$\hat X=\hat U_{\mu\nu}\,\hat q\,\hat U^{\dagger}_{\mu\nu}
=\mu \hat q+\nu\hat p$$
$$\fl
\hat P=\hat U_{\mu\nu}\,\hat p\,\hat U^{\dagger}_{\mu\nu}=
\frac{1+\sqrt{1-4\mu^2\nu^2}}{2\mu}\,\hat p-
\frac{1-\sqrt{1-4\mu^2\nu^2}}{2\nu}\,\hat q.
$$

Using the approach of \cite{MendesJPA} one can 
obtain the relationship
$$\hat U(X,\mu,\nu)=\delta(X-\mu \hat q-\nu\hat p).$$
In the case we are considering, the inverse transform 
determining the operator in terms of tomogram
[see equation~(\ref{eq.2})] will be of the form
\begin{equation}\label{eq.55}
\hat A=\int w_{\hat A}(X,\mu,\nu)
\hat D(X,\mu,\nu)\,dX\,d\mu\,d\nu
\end{equation}
where~\cite{Dariano,ManciniJMO}
\begin{equation}\label{eq.56}
\hat D({\bf x})\equiv\hat D(X,\mu,\nu)=\frac{1}{2\pi}
\exp\left(iX-i\nu\hat p-i\mu\hat q\right)
\end{equation}
i.e.
\begin{equation}\label{eq.56q}
\hat D(X,\mu,\nu)
=\frac{1}{2\pi}\exp(iX)\hat D\Big(\mbox{\boldmath$\xi$}
(\mu,\nu)\Big).
\end{equation}
The unitary displacement operator in (\ref{eq.56q}) reads now
$$
\hat D\Big(
\mbox{\boldmath$\xi$}(\mu,\nu)\Big)
=\exp\Big(\mbox{\boldmath$\xi$}(\mu,\nu)     
\hat a^+-{\mbox{\boldmath$\xi$}}^*(\mu,\nu)\hat a\Big)
$$
where $\mbox{\boldmath$\xi$}(\mu,\nu)=\xi_1+i\xi_2$
with $\xi_1=\mbox{Re}\,
(\mbox{\boldmath$\xi$})={\nu}/{\sqrt2}$
 and $\xi_2=\mbox{Im}\,(\mbox{\boldmath$\xi$})
=-{\mu}/{\sqrt2}$.

Trace of the above operator which provides the kernel
determining the trace of an arbitrary operator in the
tomographic representation reads
$$\mbox{Tr}\,\hat D({\bf x})=
e^{iX}\delta (\mu)\delta(\nu).$$ 
The creation and annihilation operators are
determined by formula (\ref{eq.29}).
The function $w_{\hat A}(X,\mu,\nu)$ satisfies the relation
\begin{equation}\label{eq.56'}
w_{\hat A}\left(\lambda X,\lambda \mu,\lambda\nu\right)
=\frac{1}{|\lambda|}\,w_{\hat A}(X,\mu,\nu).
\end{equation}
This means that the tomographic symbols of operators are homogeneous
functions of three variables.

If one takes two operators $\hat A_1$ and $\hat A_2$,
which are expressed through the corresponding functions
by the formulae
\begin{eqnarray}
\hat A_1&=&\int
w_{\hat A_1}(X',\mu',\nu')\hat D(X',\mu',\nu')\,dX'\,d\mu'
\,d\nu'
\nonumber\\
&&\label{eq.57}\\
\hat A_2&=&\int
w_{\hat A_2}(X'',\mu'',\nu'')
\hat D(X'',\mu'',\nu'')dX''\,d\mu''\,d\nu''
\nonumber
\end{eqnarray}
and $\hat A$ denotes the product of $\hat A_1$
and $\hat A_2$, then
the function $w_{\hat A}(X,\mu,\nu)$, which corresponds
to $\hat A$, is the star-product of functions 
$w_{\hat A_1}(X,\mu,\nu)$
and $w_{\hat A_2}(X,\mu,\nu)$, i.e.
$$
w_{\hat A}(X,\mu,\nu)=w_{\hat A_1}(X,\mu,\nu)
*w_{\hat A_2}(X,\mu,\nu)
$$
reads
\begin{equation}\label{eq.58}
w_{\hat A}(X,\mu,\nu)=\int w_{\hat A_1}({\bf x}'')
w_{\hat A_2}({\bf x}')K({\bf x}'',{\bf x}',
{\bf x})\,d{\bf x''}\,d{\bf x'}
\end{equation}
with kernel given by
\begin{equation}\label{eq.59}
K({\bf x}'',{\bf x}',{\bf x})=
\mbox{Tr}\left[\hat D(X'',\mu'',\nu'')
\hat D(X',\mu',\nu')\hat U(X,\mu,\nu)\right].
\end{equation}
The explicit form of the kernel reads
\begin{eqnarray}\label{KERNEL}
\fl K(X_1,\mu_1,\nu_1,X_2,\mu_2,\nu_2,X\mu,\nu)\nonumber\\
\fl =\frac{\delta\Big(\mu(\nu_1+\nu_2)-\nu(\mu_1+\mu_2)\Big)}{4\pi^2}
\,\exp\left(\frac{i}{2}\Big\{\left(\nu_1\mu_2-\nu_2\mu_1\right)
+2X_1+2X_2\right.\nonumber\\
\fl \left.\left. -\left[\frac{1-\sqrt{1-4\mu^2\nu^2}}{\nu}
\left(\nu_1+\nu_2\right)+\frac{1+\sqrt{1-4\nu^2\mu^2}}{\mu}
\left(\mu_1+\mu_2\right)
\right]X\right\}\right).
\end{eqnarray}
The kernel for the star-product of $N$ operators is
\begin{eqnarray}\label{KERNELSTAR}
\fl 
K\left(X_1,\mu_1,\nu_1,X_2,\mu_2,\nu_2,\ldots,
X_N,\mu_N,\nu_N,X,\mu,\nu\right)\nonumber\\
\fl =\frac{\delta\left(\mu\sum_{j=1}^N\nu_j-\nu\sum_{j=1}^N\mu_j\right)}
{(2\pi)^N}\,\exp\left(\frac{i}{2}\,\left\{\sum_{k<j=1}^N
\left(\nu_k\mu_j-\nu_j\mu_k\right)+2\sum_{j=1}^NX_j\right.\right.
\nonumber\\
\fl \left.\left.-\left[
\frac{1-\sqrt{1-4\mu^2\nu^2}}{\nu}
\left(\sum_{j=1}^N\nu_j\right)+
 \frac{1+\sqrt{1-4\mu^2\nu^2}}{\mu}\left(\sum_{j=1}^N\mu_j\right)
\right]X\right\}\right).
\end{eqnarray}
The above kernel can be expressed in terms of the kernel determining
the star-product of two operators.

\section{Deformed commutation relations and Poisson brackets} 

We shall consider now deformations of the associative product 
among operators.
We replace the usual product by the following  
$k$-product~\cite{SudarshanJModPhys}
\begin{equation}\label{WW18} 
(\hat A\hat B)_k
=\hat A e^{\lambda \hat k}\hat B
\end{equation} 
where $\lambda$ is a numerical parameter. 
For $\lambda=0$, the $k$-product 
(\ref{WW18}) coincides with the standard 
product of linear operators.  
The deformed commutator arising from the deformation of
the associative product will be 
\begin{equation}\label{WW19}
[\hat A, \hat B]_k=\hat A e^{\lambda\hat k}
\hat B-\hat B e^{\lambda\hat k}\hat A.
\end{equation}
This commutator defines a new Lie algebra
structure on the space of operators. 
In connection with previous consideration, 
we may introduce a deformed 
star-product ($k$ star-product or deformed Moyal product). 
We define a $k$ star-product of two functions 
in the following way:
\begin{equation}\label{WW20}
f_A({\bf x})\ast_k f_B({\bf x})=
\mbox {Tr}\left[\hat Ae^{\lambda \hat k}\hat B
\hat U({\bf x})\right]. 
\end{equation}
One can see that this deformed $k$ star-product of two symbols 
may be expressed through the usual nondeformed star-product 
\begin{equation}\label{WW48}
f_1\ast_k f_2=(f_1*f_k)*f_2
=f_1*(f_k *f_2). 
\end{equation}
Having written this new product in terms of the standard one,
the deformed Poisson brackets (or Moyal brackets) will be 
\begin{equation}\label{WW49}
\{f_1,f_2\}_k=f_1*f_k *f_2- f_2*f_k *f_1.
\end{equation} 
It is now clear that all our previous considerations can be 
repeated for this deformed product of ${\bf x}$-symbols as in 
(\ref{WW48}) and (\ref{WW49}), or as $(q,p)$ in the Wigner--Weyl 
case, or as $(X,\mu,\nu)$ in the tomographic case.

Whenever the initial dynamics has $\hat k$ as a constant of the 
motion~\cite{MarmoBregenz},
it will be compatible with the deformed $k$-product and therefore
with all subsequent considerations. Of course, more general deformations
of the associative products may be, and should be, considered if in the
classical limit we want to recover the many facets of biHamiltonian
descriptions for completely integrable systems~\cite{Carin1,Carin2}.
In the biHamiltonian description, the equation of motion can be obtained
using different Hamiltonians and different commutation relations.
This is true both in the classical and the quantum setting.
To give an insight on these possible more general commutation relations,
we consider in appendix~1 the case of 2$\times$2-matrices. The extension
to operators may be achieved by considering entries of these matrices 
to be operators, i.e. by decomposing 
${\cal H}={\cal H}_1\oplus{\cal H}_2$.

\section{Conclusions}

We summarize the main results of our paper.
We have presented the well-known scheme of the star-product procedure in
some convenient form which desribed also the case of the tomographic map.
 
The star-product procedure gives the possibility to clarify the difference
between classical and quantum dynamics.
The explicit form of the difference is expressed by 
the fact that classical dynamics is described by Hamilton equations 
for $c$-number momentum
and position, and quantum dynamics is described by
Heisenberg equations for momentum and position operators. 
Also the state of
a system in classical statistical mechanics is 
associated with a joint  
probability distribution function of position 
and momentum,
but in quantum theory the state is associated
with Hermitian nonnegative density 
operator~\cite{vonNeuman}
with matrix elements for pure states 
expressed in terms of the wave function
satisfying Schr\"odinger evolution 
equation~\cite{Schroedinger26}.
Attempts to make closer the description of the classical and 
quantum pictures have taken place
during all the period of existence of 
quantum mechanics. For the description
of quantum states, Wigner introduced~\cite{Wigner32}
the quasidistribution function on the phase space, which
has many properties similar
to classical joint probability distribution on
the classical phase space.

Nevertheless, since the
Wigner function can take negative values, it is 
obvious that this function cannot 
serve as a probability distribution, 
which must be always a nonnegative
function. One can use the same map not only for the 
density operator but also for other
quantum observables, e.g., for position 
and momentum operators and arbitrary
functions of the noncommuting positions and momenta.
Due to noncommutativity of generic quantum observables, 
the ordering of position
and momentum operators (or creation and annihilation 
operators) plays an essential
role in mapping the operator-functions 
of the position and momentum onto
$c$-number functions on the phase space.
The Wigner quasidistribution corresponds to the
symmetric ordering of the position and momentum. 
The Glauber--Sudarshan and Husimi
quasidistributions correspond to antinormal 
and normal ordering of creation and
annihilation operators, respectively.
The one-parametric family of
quasidistributions describing the $s$-ordering 
of the operators was introduced by Cahill and 
Glauber~\cite{CahillGlauber67}.

For the values of the parameters $s=0,1,-1$, 
the $s$-quasidistributions provide the Wigner,
Glauber--Sudarshan and Husimi quasidistributions,
respectively. Thus, one has different maps from
operators (not only density operators) onto
functions on the phase space,
and the map properties depend on the continuous
parameter $s$.

We have presented a unified approach to construct 
all these different invertible maps from operators acting 
on a Hilbert space
onto functions (symbols) of several variables. The construction
can be extended to consider also maps from operators onto
functions of infinite number of variables (functionals). 
We have established invertible relations between symbols 
of different sorts. 

We have embedded the tomographic map
into the presented general scheme and studied different 
star-products of the functions corresponding to different 
maps and calculated the kernels of the integral operators 
which define star-product. 
The importance of this result is related to the fact that the
tomographic map created a new formulation of quantum mechanics 
in which the standard probability density describes the quantum
state instead of the wave function and density matrix.
The results of this paper demonstrate that the new formulation
of quantum mechanics can be given in terms of the well-known
procedure of  star-product quantization but with a specific kernel
which was not known till now.
The explicit form of the kernel
for star-product of tomograms and $s$-ordered symbols 
is a new contribution of this paper. 
Deformations of star-product of operator-symbols are 
also considered.  

\section*{Acknowledgments}

Olga~V~M and V~I~M thank Dipartimento di
Scienze Fisiche, Universit\'a ``Federico~II'' 
di Napoli and Istitito Nazionale di Fisica Nucleare, 
Sezione di Napoli for kind hospitality, the Russian 
Foundation for Basic Research for partial support 
under Projects~Nos.~00-02-16516 and 99-02-17753 
and the Ministry for Industry, Sciences and Technology of 
the Russian Federation for the support within the framework
of the Programs ``Optics. Laser Physics'' and ``Fundamental
Nuclear Physics.'' This paper has been supported by PRIN SINTESI.

\section*{Appendix~1.~Associative products on vector spaces of 
finite dimensions}

Below we discuss some possible associative products on
$n$$\times$$n$-matrices which differ from the standard one.
Let us consider the simplest example of 2$\times$2-matrices
$$\pmatrix{a_{11}&a_{12}\cr a_{21}&a_{22}}
\qquad
\pmatrix{b_{11}&b_{12}\cr b_{21}&b_{22}}.
$$
The set of all 2$\times$2-matrices can be mapped onto the set 
of 4-vectors in the four-dimensional linear space by means of the 
invertible correspondence rule
\begin{equation}\label{ap2}
\fl 
a\leftrightarrow {\bf A}=\left(a_{11},a_{12},a_{21},a_{22}\right)
\qquad
b\leftrightarrow {\bf B}=\left(b_{11},b_{12},b_{21},b_{22}\right).
\end{equation}
Due to this, the product of matrices $a$ and $b$
can be considered 
as a product of the corresponding vectors.
We define the product of two vectors as the bilinear function
$$
{\bf C}\left({\bf A},{\bf B}\right)={\bf A}\odot{\bf B}
$$
such that 
$$
\left[C\left(A,B\right)\right]_k=M_k^{mn}A_mB_n
$$
or
\begin{equation}\label{ap3}
C_k=\sum_{n,s=1}^4 A_nM^{ns}_kB_s\qquad n,s,k=1,2,3,4.
\end{equation}
The associativity condition 
$$
\left({\bf A}\odot{\bf B}\right)\odot{\bf C}={\bf A}\odot
\left({\bf B}\odot {\bf C}\right)
$$
imposes the following equation 
\begin{equation}\label{ap5}
\sum_{m=1}^4 M^{nm}_lM^{sk}_m=
\sum_{m=1}^4 M^{ns}_mM^{mk}_l.
\end{equation}
All the solutions of equation~(\ref{ap5}) provide all possible
associative products on the space of 
2$\times$2-matrices. For example, four matrices
\begin{eqnarray*}
\fl M_1=\pmatrix{1&0&0&0\cr 0&0&0&0\cr 0&0&0&0\cr 0&0&0&0}
\qquad
M_2=\pmatrix{0&0&0&0\cr 0&1&0&0\cr 0&0&0&0\cr 0&0&0&0}
\nonumber\\
\fl M_3=\pmatrix{0&0&0&0\cr 0&0&0&0\cr 0&0&1&0\cr 0&0&0&0}
\qquad
M_4=\pmatrix{0&0&0&1\cr 0&0&0&0\cr 0&0&0&0\cr 0&0&1&0}
\nonumber
\end{eqnarray*}
which satisfy equation~(\ref{ap5}), provide the 
following associative product 
of 2$\times$2-matrices of the form
$$
a\odot b=\pmatrix{a_{11}b_{11}&a_{12}b_{12}\cr
a_{21}b_{21}&a_{11}b_{22}+a_{22}b_{21}}.
$$
Another solution of equation~(\ref{ap5}) of the form
\begin{eqnarray*}
\fl M_1=\pmatrix{k_{11}&0&k_{12}&0\cr k_{21}&0&k_{22}&0\cr
 0&0&0&0\cr 0&0&0&0}\qquad
M_2=\pmatrix{0&k_{11}&0&k_{12}\cr 0&k_{21}&0&k_{22}\cr 
0&0&0&0\cr 0&0&0&0}\nonumber\\
\fl M_3=\pmatrix{0&0&0&0\cr 0&0&0&0\cr k_{11}&0&k_{12}&0\cr 
k_{21}&0&k_{22}&0}\qquad
M_4=\pmatrix{0&0&0&0\cr 0&0&0&0\cr 0&k_{11}&0&k_{12}\cr 
0&k_{21}&0&k_{22}}\nonumber
\end{eqnarray*}
 provides the following deformed associative product 
discussed in~\cite{MarmoBregenz}
\begin{equation}\label{ap9}
a\odot b=akb
\end{equation}
where the product of three matrices in the right-hand side
of equation~(\ref{ap9}) is the standard one
$$
k=\pmatrix{k_{11}&k_{12}\cr k_{21}&k_{22}}.
$$
If the matrices $M_k$ are symmetric ones, the product of
the matrices is commutative. 
One can compare equation~(\ref{ap5}) with the Jacobi identity 
for structure constants of a Lie group $C_{sk}^{(m)}$
\begin{equation}\label{ap11}
\sum_m C_{sk}^{(m)}C_{nm}^{(l)}+
\sum_m C_{kn}^{(m)}C_{sm}^{(l)}+
\sum_m C_{ns}^{(m)}C_{km}^{(l)}=0\,.
\end{equation}
Equation~(\ref{ap11}) is also quadratic relation 
similar to equation~(\ref{ap5}).
The standard associative product of 
matrices may be defined not only for square
matrices but also for the rectangular ones. 
It is possible to provide 
different associative products also for  
rectangular matrices if one maps the 
matrices onto the set of vectors in a linear space. 
The dimensionality of the linear space is the higest dimensionality
of the square matrices involved in the product. Thus, the problem
of different associative products for rectangular matrices may be reduced 
to the problem of different associative products of square matrices 
under discussion. The defined associative product of 2$\times$2-matrices
can be extended easily to $n$$\times$$n$-matrices and to the products
of kernels of the operators in infinite-dimensional spaces.

\section*{Appendix~2.~Associative product on vector spaces 
of infinite dimensions}

We will present the corresponding formulae for the associative
product of functions $f({\bf x})$, where {\bf x}$=(x_1,x_2,\ldots,x_n)$.
The function $f({\bf x})$ can be considered as a set of vector 
components $f_{\bf x}\equiv f({\bf x}).$ Due to this, one can use the
procedure presented for 2$\times$2-matrices. The standard product
of functions                     
\begin{equation}\label{P1}
(f_1f_2)({\bf x})=f_1({\bf x})f_2({\bf x})
\end{equation}                       
can be presented in the integral form          
\begin{equation}\label{P2}
f_1({\bf x})f_2({\bf x})=\int f_1({\bf y})f_2({\bf z})    
K({\bf x},{\bf y},{\bf z})\,d{\bf y}\,d{\bf z}
\end{equation}                                 
where the kernel has the form            
\begin{equation}\label{P3}
K({\bf x},{\bf y},{\bf z})=\delta({\bf x}-{\bf y})
\,\delta({\bf x}-{\bf z})
\end{equation}         
to reproduce the point-wise product.
The product~(\ref{P1}) is known to be associative.
Introducing the following notation for the kernel 
$$
M^{\bf yz}_{\bf x}\equiv K({\bf x},{\bf y},{\bf z})
$$               
one sees that the kernel in (\ref{P2}) is completely 
analogous to $M^{ns}_k$ in equation~(\ref{ap3}). It is now 
clear that instead of the kernel (\ref{P3}) one can use other kernels,
and define other associative star-product for the functions 
$f_1({\bf x})$ and $f_2({\bf x})$ by setting
\begin{equation}\label{P4}
\left(f_1*f_2\right)({\bf x})=\int
f_1({\bf y})f_2({\bf z})    
K({\bf x},{\bf y},{\bf z})\,d{\bf y}\,d{\bf z}.
\end{equation}                                 
The associativity condition requires that the integral equation 
for the kernel 
\begin{equation}\label{P45}
\int K({\bf x},{\bf y}, {\bf z})K({\bf z},{\bf l}, {\bf t})
\,d{\bf z}=\int K({\bf x},{\bf z}, {\bf t})
K({\bf z},{\bf y}, {\bf l})\,d{\bf z}
\end{equation}            
be satisfied.
If $K({\bf x},{\bf y}, {\bf z})= 
K({\bf x},{\bf z}, {\bf y})$,
the star-product of functions turns out to be a commutative product.
The general form for the solution of the associativity condition was
discussed in \cite{Tyutin1} and for grassmanian variables 
in \cite{Tyutin2}.

\end{document}